\documentclass[12pt]{article}

\usepackage{cite}
\usepackage{epsfig}
\usepackage{amsmath}
\usepackage{array}
\usepackage{pstricks}
\usepackage{bbold}

\def\mathswitch#1{\relax\ifmmode#1\else$#1$\fi}
\def\mathswitchr#1{\relax\ifmmode{\mathrm{#1}}\else$\mathrm{#1}$\fi}

\newcommand{\anc}{\rule{0mm}{0mm}}

\newcommand{\mycaption}[1]{\caption{\sl #1}}

\begin{document}
\thispagestyle{empty}

\def\thefootnote{\fnsymbol{footnote}}

\begin{flushright}
\end{flushright}

\vspace{1cm}

\begin{center}

{\Large\sc {\bf On the Numerical Evaluation of Loop Integrals\\[.5ex]
 With Mellin-Barnes Representations}}
\\[3.5em]
{\large\sc
Ayres~Freitas, Yi-Cheng~Huang
}

\vspace*{1cm}

{\sl
Department of Physics \& Astronomy, University of Pittsburgh,\\
3941 O'Hara St, Pittsburgh, PA 15260, USA
}

\end{center}

\vspace*{2.5cm}

\begin{abstract}
An improved method is presented for the numerical evaluation of multi-loop
integrals in dimensional regularization. The technique is based on Mellin-Barnes
representations, which have been used earlier to develop algorithms for the
extraction of ultraviolet and infrared divergencies. The coefficients of these
singularities and the non-singular part can be integrated numerically. However, 
the numerical integration often does not converge for diagrams with massive
propagators and physical branch cuts. In this work, several steps are proposed
which substantially improve the behavior of the numerical integrals.  The
efficacy of the method is demonstrated by calculating several two-loop examples,
some of which have not been known before.
\end{abstract}

\setcounter{page}{0}
\setcounter{footnote}{0}

\newpage


\section{Introduction}

Perturbation theory is one of the most important tools for the calculation  of
particle physics observables. In many cases, radiative corrections are important
to match the experimental precision. General and efficient techniques for the
computation of one-loop corrections have been developed over many years,
resulting in automated frameworks for the calculation of processes with up to
six external legs \cite{onesix}. Most methods for one-loop corrections reduce
the loop integrals to a well-defined set of master integrals by
Passarino-Veltman and Melrose reduction \cite{pave}, integration-by-parts and
Lorentz-invariance relations \cite{ibp}, or on-shell cut techniques
\cite{onshell}. See Ref.~\cite{onerev} and references therein for a more
detailed overview.

In some cases, however, one-loop corrections are not sufficient and two- or even
three-loop corrections becomes relevant. Typical examples are electroweak
precision observables \cite{ewprec,ew2} and production of gauge bosons
\cite{lhcw}, top quarks \cite{lhct,lhct2} or Higgs bosons \cite{lhch} at the
Large Hadron Collider (LHC). Many highly complex calculations have been
performed for this purpose by several authors (see references in previous
sentence), involving specialized techniques tailored to each particular process.
For two-loop amplitudes with a small number of different mass and momentum
scales one can follow the traditional approach of reduction to a small number of
master integrals, which are then calculated analytically. However, this strategy
does not work for applications with a large number of different scales, due to
the complexity of the expressions and the fact that in general the master
integrals cannot be determined analytically beyond the one-loop level.

Fully numerical techniques for the evaluation of two- and higher-loop integrals
face two main challenges: extraction of ultraviolet (UV) and infrared and
collinear (IR) singularities, as well as stability  and efficiency of the
numerical integrations. Methods based on dispersion relations \cite{disp,ew2}
and integrations over Feynman parameters \cite{feynp} result in low-dimensional
numerical integrals with good convergence properties, but they do not offer a
general algorithm for the isolation of divergencies. Two powerful approaches for
a straightforward extraction of $1/(4-D)$ poles in dimensional regularization
are sector decomposition \cite{sec} and Mellin-Barnes representations
\cite{mb,mb2,mellinbarnes}. However, these methods lead to complicated
multi-dimensional numerical integrals, which have convergence problems for
diagrams with physical branch cuts. For sector decomposition, the instability
related to physical branch cuts can be circumvented by the integration contour
deformation proposed in Ref.~\cite{ns}.

This article instead focuses on the numerical integration of Mellin-Barnes (MB)
representations. Compared to sector decomposition, MB representations of Feynman
integrals have the advantage that they typically lead to substantially shorter
expressions in the integrand. However, MB representation will only become a
viable general tool for numerical two-loop integrations if the behavior of the
numerical integrals can be substantially improved. In this paper, several
techniques for this purpose are presented, which can also be used in combination
to optimize the convergence properties of the integrals. In particular, complex
variable transformations and relations that reduce the number of integration
dimensions prove to be very useful.

Section~\ref{sc:mb} starts out by reviewing the employment of MB representations
for loop integrals and explaining the origin of the numerical instabilities in
the MB integrals. In section~\ref{sc:num} the new methods to deal with these
instabilities are described in detail. Numerical results for several examples of
two-loop integrals are presented in section~\ref{sc:ex}. Where applicable, they
are compared to existing results in the literature, which have been obtained by
other methods. In addition, results are shown for some integrals appearing in
the calculation of two-loop QCD corrections to $gg\to t\bar{t}$ which have not
been calculated before. Finally, the main findings are summarized in
section~\ref{sc:sum}.


\section{Mellin-Barnes representations}
\label{sc:mb}


The following discussion of the construction of MB representations and the
isolation of UV and IR singularities is largely based on Ref.~\cite{mb}. 
Let us start with a one-loop $n$-point integral
\begin{equation}
I_1 = \int \frac{d^Dq}{i\pi^{D/2}} \, \Bigl [
[q^2-m_0^2]^{\mu_0} 
[(q+p_1)^2-m_1^2]^{\mu_1} \cdots 
[(q+p_1+p_2+\ldots+p_n)^2-m_n^2]^{\mu_n} 
\Bigr ]^{-1}
\end{equation}
and its Feynman parametrization
\begin{equation}
\begin{aligned} I_1 =
(-1)^M &\frac{\Gamma(M-D/2)}{\Gamma(\nu_0)\cdots\Gamma(\nu_n)}
\int_0^1 dx_0\cdots dx_n \; \delta(1-x_0 - \ldots -x_n) \\
&\times \frac{x_0^{\mu_0-1}\cdots x_n^{\mu_n-1}}{\left[
-\sum_{i,j=1}^n (p_i-p_j)^2 x_i x_j - \sum_{i=1}^n (p_i^2 x_i^2 - m_i x_i) + m_0
x_0 - i\epsilon \right]^{M-D/2}},
\end{aligned}
\end{equation}
where $M=\mu_0+\dots+\mu_n$. The integrand denominator 
can be transformed into a MB representation by
\begin{equation}
\begin{aligned}
\frac{1}{(A_0+\ldots+A_m)^Z} = \frac{1}{(2\pi i)^m} \int_{{\cal C}_1} dz_1
\cdots &\int_{{\cal C}_m} dz_m \; A_1^{z_1} \cdots A_m^{z_m}
A_0^{-Z-z_1-\ldots-z_m} \\
&\times \frac{\Gamma(-z_1) \cdots \Gamma(-z_m)\Gamma(Z+z_1+\ldots+z_m)}{\Gamma(Z)}
\end{aligned}
\end{equation}
where the following two conditions must be met:
\begin{enumerate}
\item
The integration contours ${\cal C}_i$ for $z_i$ are straight lines parallel to
the imaginary axis chosen such that all arguments of the gamma functions have
positive real parts.
\item
The $A_i$ are complex numbers with $|\arg(A_i)-\arg(A_j)|<\pi$ for any pair
$i,j$. \label{arg}
\end{enumerate}
Satisfying the second condition potentially requires some adjustments of the
Feynman parametrization. For example, the scalar diagram in Fig.~\ref{fig:top1} is
represented by the Feynman integral
\begin{equation}
I_1^{\rm fig \ref{fig:top1}} = \Gamma(4-\tfrac{D}{2})\int_0^1 dx_0dx_1dx_2dx_3 \,
\frac{\delta(1-x_0-x_1-x_2-x_3)}{\left[
 -sx_1x_3-tx_0x_2+m^2x_2-m^2x_1x_2-m^2x_2x_3-i\epsilon\right]^{4-\frac{D}{2}}},
\label{exa}
\end{equation}
where the terms with $m^2$ violate condition \ref{arg} above.
More generally, problems with the application of the MB representation arise if 
one mass or momentum parameter appears both with positive and negative signs in
the denominator of the Feynman integral. This complication
can be avoided by using the relation $\sum_ix_i=1$. In the case of
eq.~\eqref{exa}, this relation allows to rewrite it as
\begin{equation}
I_1^{\rm fig \ref{fig:top1}} = \Gamma(4-\tfrac{D}{2})\int_0^1 dx_0dx_1dx_2dx_3 \,
\frac{\delta(1-x_0-x_1-x_2-x_3)}{\left[
 -sx_1x_3-tx_0x_2+m^2x_0x_2+m^2x_2^2-i\epsilon\right]^{4-\frac{D}{2}}},
\label{exa2}
\end{equation}
so that all terms with $m^2$ now have positive signs.
The term $-sx_1x_3$ does not lead to any difficulties since one can imagine to
temporarily shift $s$ slightly off the real axis into the complex plane so that
$|arg(-sx_1x_3)| < \pi$ is satisfied. Thus the Mellin-Barnes formula can be
safely applied to eq.~\eqref{exa2}.
\begin{figure}
\vspace{1ex}
\begin{center}
\rule{0mm}{0mm}%
\psline(-1,0)(1,0)
\psline(-1,-2)(1,-2)
\psline(-2.2,0)(-1,0)(-1,-2)(-2.2,-2)
\psline[linewidth=2pt](2.2,0)(1,0)(1,-2)(2.2,-2)
\psdot[dotscale=1.5](-1,0)
\psdot[dotscale=1.5](1,0)
\psdot[dotscale=1.5](-1,-2)
\psdot[dotscale=1.5](1,-2)
\end{center}
\vspace{1.7cm}
\mycaption{Scalar box diagram corresponding to eq.~\eqref{exa}. Thick lines
indicate propagators with mass $m$, while thin lines are massless, and external
legs are constrained to be on-shell.
\label{fig:top1}} 
\end{figure}

After the introduction of the MB representation, the integration over the
Feynman parameters can be carried out straightforwardly, using
\begin{equation}
\int_0^1 dx_0 \cdots dx_n \; \delta(1-x_0 - \ldots -x_n) \,
x_0^{\alpha_0-1} \cdots x_n^{\alpha_n-1}
= \frac{\Gamma(\alpha_0) \cdots \Gamma(\alpha_n)}{\Gamma(\alpha_0+\ldots+\alpha_n)}
,
\end{equation}
assuming that the exponents satisfy
\begin{equation}
{\rm Re}(\alpha_i) > 0.
\end{equation}

A MB representation for a two-loop integral can be obtained by first deriving a
MB representation for one sub-loop, as described above, and inserting the result
into the second loop, which then can be transformed into a MB representation
with the same procedure. In general, the MB representation for the first
sub-loop will contain invariants depending on the second loop momentum, which
can be interpreted as new propagators for the second integration.

\begin{figure}
\vspace{1ex}
\begin{center}
\rule{0mm}{0mm}%
\psline(-2,-1)(2,-1)
\pscircle(0,-1){1}
\psdot[dotscale=1.5](-1,-1)
\psdot[dotscale=1.5](1,-1)
\rput[t](0,-0.2){$m_1$}
\rput[t](0,-2.1){$m_3$}
\rput[t](0,-1.2){$m_2$}
\rput[lb](-2,-0.9){$\displaystyle p\atop\longrightarrow$}
\end{center}
\vspace{1.7cm}
\mycaption{``Sunset'' two-loop diagram. As before, all propagators are
assumed to be scalars.
\label{fig:sunset}} 
\end{figure}
As an example, consider the ``sunset'' diagram in Fig.~\ref{fig:sunset}. The MB
representation for the upper sub-loop is given by
\begin{align}
I_1^{\rm fig \ref{fig:sunset},sub} = \frac{1}{(2\pi i)^2} \int dz_1dz_2 \, 
&
(m_1^2)^{-\varepsilon-z_1-z_2} (m_2^2)^{z_2} (-q_2^2)^{z_1} \;
\Gamma(-z_1) \Gamma(-z_2) \Gamma(1+z_1+z_2) \nonumber \\ 
&\times\frac{\Gamma(1-\varepsilon-z_2)
\Gamma(\varepsilon+z_1+z_2)}{\Gamma(2-\varepsilon+z_1)},
\end{align}
where $\varepsilon=(4-D)/2$ and $q_2$ is the internal momentum of the second
(lower) loop. For the second integration one then obtains
\begin{align}
I_2^{\rm fig \ref{fig:sunset}} &= \frac{1}{(2\pi i)^2} 
\int dz_1dz_2 \, 
\int \frac{d^Dq_2}{i\pi^{D/2}} \,
\Bigl [
[q^2]^{-z_1} 
[(q_2+p)^2-m_3^2] 
\Bigr ]^{-1} \nonumber \\ 
\times (-&1)^{z_1} 
(m_1^2)^{-\varepsilon-z_1-z_2} (m_2^2)^{z_2} \;
\Gamma(-z_1) \Gamma(-z_2) \Gamma(1+z_1+z_2) 
\frac{\Gamma(1-\varepsilon-z_2)
\Gamma(\varepsilon+z_1+z_2)}{\Gamma(2-\varepsilon+z_1)},
\end{align}
so that the $(q_2^2)^{z_1}$ gives rise to a new propagator. After constructing
the MB representation for the second loop, the final result is
\begin{align}
I_2^{\rm fig \ref{fig:sunset}} = &\;\frac{-1}{(2\pi i)^3} \int dz_1dz_2dz_3 \, 
(m_1^2)^{-\varepsilon-z_1-z_2} (m_2^2)^{z_2} (m_3^2)^{1-\varepsilon+z_1-z_3} 
(-p^2)^{z_3} \; \Gamma(-z_2) \Gamma(-z_3) 
\nonumber\\  &\;
\times \Gamma(1+z_1+z_2) \Gamma(z_3-z_1) 
\frac{\Gamma(1-\varepsilon-z_2)\Gamma(\varepsilon+z_1+z_2)\Gamma(\varepsilon-1-z_1+z_3)}
{\Gamma(2-\varepsilon+z_3)}.
\label{ss}
\end{align}
The MB integral can then be be performed numerically.

As mentioned above, the Mellin-Barnes formula and the Feynman parameter
integration require the arguments of all gamma functions to have a positive real
part. In general, this requires $\varepsilon$ to differ from zero by a finite
amount. When taking the limit $\varepsilon\to 0$, the poles of some gamma
functions can cross some of the integration contours, so that one needs to
correct for this by picking up the residue of the integrand for crossed each
pole. As a result, for $\varepsilon\to 0$ one ends up with the original integral
plus a sum of terms with lower integration dimension stemming from these
residues. An algorithm for this procedure is described in Refs.~\cite{mb,mb2}.

Since for appropriate choices of the contours each integral in this sum is
regular, the $1/\varepsilon$ singularities must be contained in the coefficients
of these integrals. Therefore this method allows to extract all UV and IR
divergencies in a systematic way.


\section{Improving numerical convergence}
\label{sc:num}


The contour integrations in the MB representation can be parametrized as
\begin{equation}
\int_{{\cal C}_i} dz_i \, f(z_i) = i \int_{-\infty}^\infty dy_i \, f(c_i+iy_i),
\label{std}
\end{equation}
where the $c_i$ are real constants. The integrand contains powers of the  kinematic
and mass invariants as well as gamma functions. 

Let us return to the example of the sunset diagram, eq.~\eqref{ss}.
For Euclidean momentum, $p^2<0$, the dependence of the
invariants on the integration variables $y_i$ is given by simple oscillating
exponentials, $e.\,g.$ $(m^2_2)^{iy_2}$, while the gamma functions rapidly tend
to zero when their arguments take values away from the real axis.
Therefore, in this case, the integrals are well behaved for numerical
integration, and a moderate finite integration interval $y_i \in [-10,
10]$ is sufficient to reach a high precision for the result.

\begin{figure}
\epsfig{figure=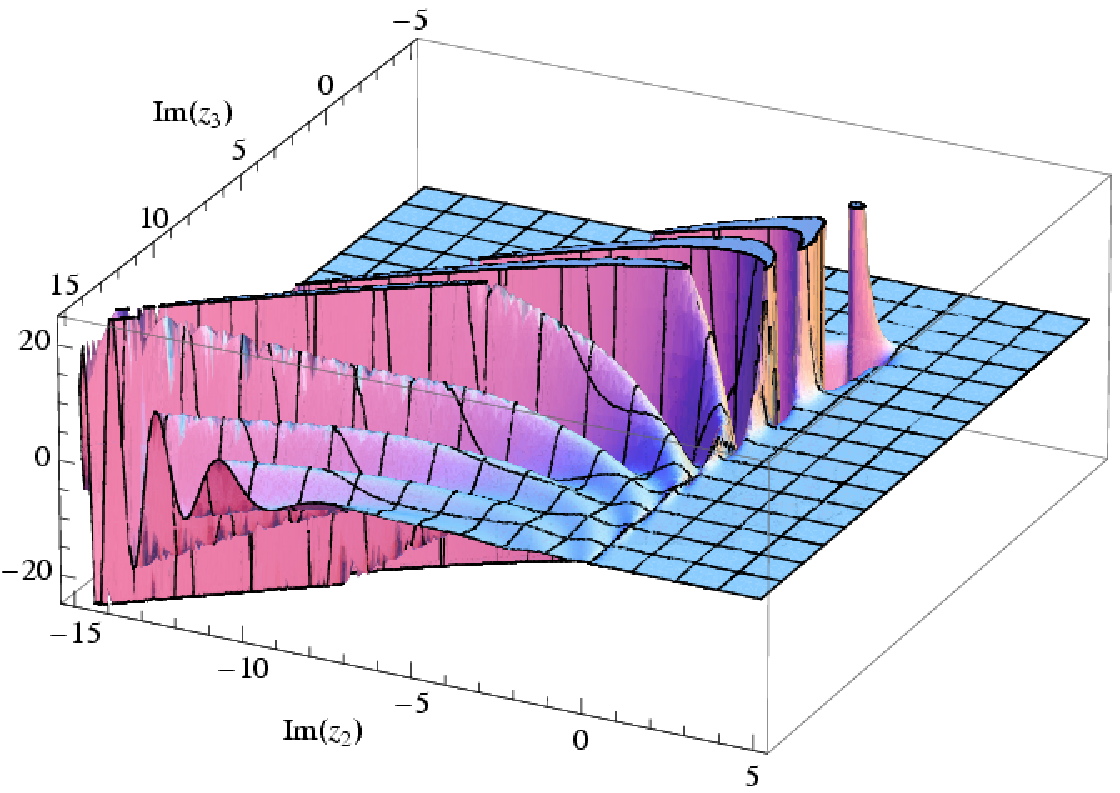, width=8cm}
\hfill
\epsfig{figure=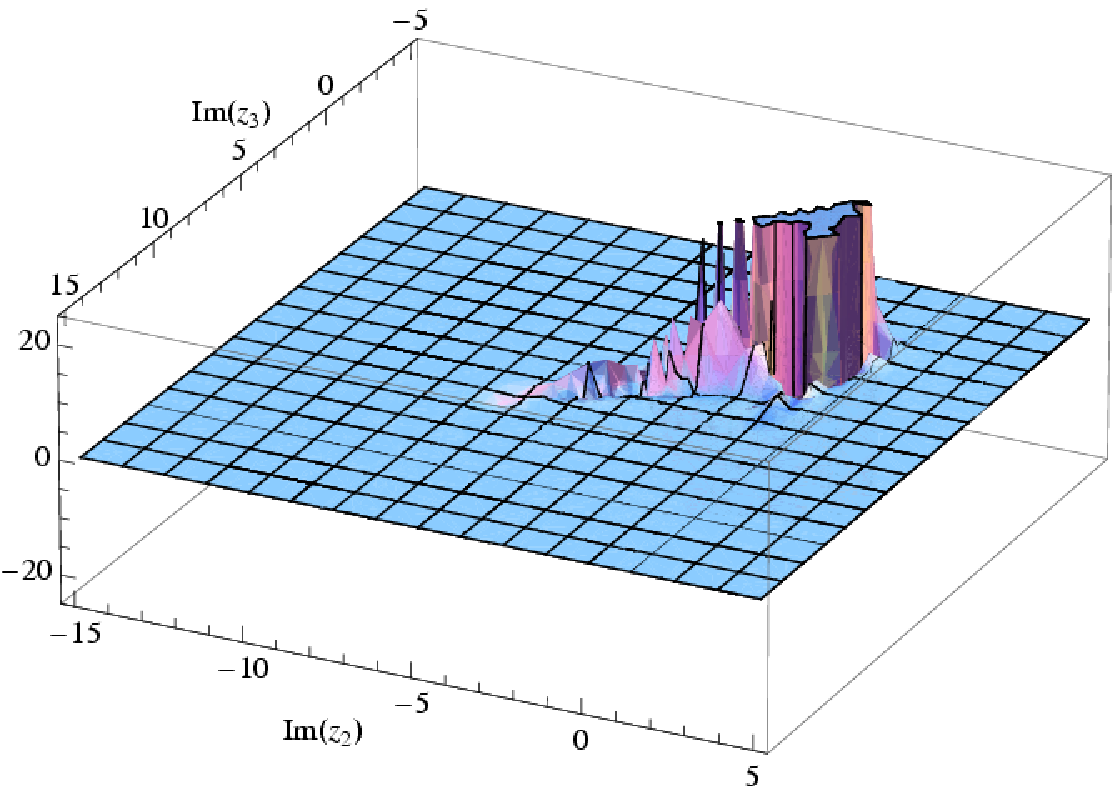, width=8cm}
\vspace{-1ex}
\mycaption{The real part of the integrand for the ``sunset'' diagram in
eq.~(\ref{ss}), for $p^2=\,$1, $m_1^2=\,$1, $m_2^2=\,$4, $m_3^2=\,$5. For the
\textbf{left panel}, the integration contours have been chosen as straight lines
parallel to the imaginary axis, and the $z_1$ integration has already been
carried out. For the \textbf{right panel}, the contours have been deformed by a
rotation in the complex plane, corresponding to $\theta=0.4$ in eq.~(\ref{expt}).
\label{fig:osc}} 
\end{figure}

However, for physical momentum, $p^2>0$, an exponentially growing term appears in the
integrand:
\begin{equation} 
(-p^2)^{z_3} = (p^2)^{c_3+iy_3} (-1-i\epsilon)^{c_3+iy_3}
	     = (p^2)^{c_3+iy_3} e^{-i\pi c_3} e^{\pi y_3}. \label{exp}
\end{equation}
Thus the integrand is an oscillating function with slowly decreasing
amplitude for $y_3\to\infty$. While the
integral is formally convergent, numerical integration routines have
difficulties dealing with this situation, since one needs to integrate to very
large values of $y_3$, $i.\,e.$ over many oscillation periods, to reach
acceptable precision. See Fig.~\ref{fig:osc} for illustration.

The problem of slowly converging oscillating tails of the integrand is generic
for any MB representation of a loop integral with physical (Minkowski) momenta.
Special techniques for numerical integration of oscillating
integrands like Filon's method \cite{filon} fail since the oscillation period
varies substantially within the integration range. Instead, as shown in the
following, suitable variable transformations can substantially improve the
convergence properties.



Next, one can exploit the freedom to deform the integration contours in the
complex plane, as long as no contour crosses a pole of the integrand. The latter
condition is rather restrictive, but if all contours are rotated in the
complex plane by the same
angle it is guaranteed that no pole crossing occurs:
\begin{equation}
c_i+iy_i \to c_i + (\theta+i)y_i. \label{deform}
\end{equation}
In eq.~\eqref{exp} this leads to
\begin{equation}
(-p^2)^{z_3} = (p^2)^{c_3+iy_3} e^{-i\pi (c_3+\theta y_i)} 
	       e^{(\pi+\theta\log p^2) y_3}. \label{expt}
\end{equation}
so that the problematic exponential $\exp(\pi y_3)$ can, in principle, be
compensated by the a suitable choice of $\theta$. For the example shown in
Fig.~\ref{fig:osc}~(right), it is evident that with this contour deformation the
integrand drops off very fast away from the origin, and thus is much better
suited for numerical integration.

For a multi-dimensional MB
integral, the choice of $\theta$ is more delicate since this one parameter
affects all integration variables. It helps to replace the variables
$y_{1,\dots,n}$
by hyper-spherical coordinates, $i.\,e.$ $n-1$ angles and one radial coordinate:
\begin{equation}
\begin{aligned}
y_1 &= r \cos \phi_1, \\
y_2 &= r \sin \phi_1 \cos \phi_2, \\
\vdots \; &\qquad\qquad \vdots \\
y_{n-1} &= r \sin \phi_1 \cdots \sin\phi_{n-2} \cos \phi_{n-1},\\
y_n &= r \sin \phi_1 \cdots \sin\phi_{n-2} \sin \phi_{n-1}.
\end{aligned}
\end{equation}
In this representation, only $r$ is affected by the modification \eqref{deform}
of the integration contours, so that one can in principle choose $\theta$
such that all problematic exponentials of the form \eqref{exp} are cancelled, 
as a function of the $\phi_i$.

After applying the variable transformations, the multi-dimensional MB integral
is sufficiently well-behaved that it can be evaluated with standard Monte-Carlo
integration algorithms like \textsc{Vegas} \cite{vegas}\footnote{For relatively simple
integrals with four or less integration dimensions, deterministic integration
methods like Gaussian quadrature prove more efficient.}. In particular, the
Monte-Carlo error decreases systematically with the number of integration
points, which is an important requirement for the convergence of the numerical
integration routine.

However, MB integrals with high dimensionality can still require a large amount
of computing time to reach a result with acceptable precision. For two-loop
diagrams one can obtain MB integrals with more than 10 dimensions, depending on
the topology and the number of independent masses and momenta involved.
Fortunately, it turns out that in most cases some of the integrations can be
performed analytically. For this purpose, the convolution theorem for Mellin
transforms proves to be useful. The Mellin transform of a function $f$ is
defined as
\begin{equation}
{\cal M}[f(x);z] = \int_0^\infty dx\,f(x) x^{z-1}.
\end{equation}
For the Mellin transform of a product of two functions, the convolution theorem 
then states
\begin{equation} 
{\cal M}[f(x) g(x);s] = \frac{1}{2\pi i}\int_{c-i\infty}^{c+i\infty}
dz \, {\cal M}[f(x);z] \, {\cal M}[g(x);s-z],
\label{mellin}
\end{equation}
where the integration is parallel to the imaginary axis and 
the integrand in eq.~\eqref{mellin} must be analytic for
Re$(z)$ in some region around $c$.

With suitable choices of $f$ and $g$
one can derive formulas for MB integrals of two or four gamma
functions:
\begin{align}
&\frac{1}{2\pi i} \int_{c-i\infty}^{c+i\infty} dz \, \Gamma(-z) \Gamma(\beta+z)
\, t^z
 = \Gamma(\beta) \,(1+t)^{-\beta} &&\hspace{-1em}\bigl [
 0 > c > 1-\text{Re}(\beta) \bigr] \label{mba} \\[1ex]
&\frac{1}{2\pi i} \int_{c-i\infty}^{c+i\infty} dz \, \Gamma(z)\Gamma(\alpha-z)\Gamma(\beta-z)
 \Gamma(1-\gamma+z) \label{mbb} \\
& \quad = \frac{\Gamma(\alpha)\Gamma(\beta)\Gamma(\alpha-\gamma+1)
 \Gamma(\beta-\gamma+1)}{\Gamma(\alpha+\beta-\gamma+1)}
&&\hspace{-1em}\bigl [
\text{Re}(\alpha),\,\text{Re}(\beta) > c > 0,\,\text{Re}(\gamma)-1
\bigr] \nonumber \\[1ex]
&\frac{1}{2\pi i} \int_{c-i\infty}^{c+i\infty} dz \, \Gamma(-z)\Gamma(\alpha+z)\Gamma(\beta+z)
 \Gamma(\gamma-\alpha-z)\,t^{-\alpha-z} \hspace{-.5em} \label{mbc} \\
& \quad = \frac{\Gamma(\alpha)\Gamma(\beta)\Gamma(\gamma)
 \Gamma(\beta+\gamma-\alpha)}{\Gamma(\beta+\gamma)} 
 \, {}_2F_1(\alpha,\gamma,\beta+\gamma,1-t)\hspace{-3em}
&&\;\;\bigl [ 0 > c > 1-\text{Re}(\beta),\,1-\text{Re}(\gamma)
\bigr]  \nonumber
\end{align}
where in each case the contours are assumed to be straight lines parallel to the
imaginary axis satisfying the conditions in square brackets, 
and ${}_2F_1$ is the Gauss hypergeometric function. These
formulas should be applied before the variable transformations. The maximal
reduction of the integration dimension is achieved if the formulas are used both
before and after the extraction of the $1/\varepsilon$ poles, since in each case one
obtains different combinations of terms that match the left-hand sides of
eqs.~\eqref{mba}--\eqref{mbc}. The reduction procedure has been implemented in a computer
algebra program based on \textsc{Mathematica},
which automatically looks for all possible applications of these equations.


\section{Examples}
\label{sc:ex}

In this section numerical results for some examples of scalar two-loop integrals
are presented. The generation of the MB representation, the extraction of the
$1/\varepsilon$ poles, and the application of the reduction formulas, as
described in the previous section, has been implemented in
\textsc{Mathematica}.  For the numerical integration a C code has been
developed, which uses Gaussian quadrature for low dimensionality and the
\textsc{Vegas} Monte-Carlo routine for high dimensionality.

In order to test the method and its computer implementation, results will be
shown first for several known
two-loop integrals, which have been calculated earlier with other methods.
A few examples that cannot be computed with any previously
published method will be presented at the end of this section.

It should be stressed that all examples in this section cannot be evaluated
numerically with MB representations by using the standard representation
eq.~\eqref{std}, as implemented $e.\,g.$ in the \textsc{mb} package of
Ref.~\cite{mb2}, see also Ref.~\cite{mb}. The reason is that for physical
momenta, $p_i>0$, the integrands have long oscillating tails that converge too
slowly. Therefore, for all cases shown in the following, the variable transformations and contour
deformations introduced in the previous sections have been used.

\paragraph{Self-energy topologies:} The scalar sunset diagram in
Fig.~\ref{fig:sunset} leads to a three-dimensional MB integral, as shown in
section~\ref{sc:mb}, together with one- and two-dimensional integrals
originating from pole crossings when the limit $\varepsilon\to 0$ is taken.
Expanding in powers of $\varepsilon$, it turns out that 
the UV poles are given by terms without integrals:
\begin{equation}
I_2^{\rm fig \ref{fig:sunset}} = \frac{m_1^2+m_2^2+m_3^2}{2\varepsilon^2}
 + \frac{1}{\varepsilon}\biggl [\sum_{i=1}^3 m_i^2(\tfrac{3}{2}-\gamma_E- \log m_i^2)
 -\frac{p^2}{4} \biggr] + I_{2,\rm fin}^{\rm fig \ref{fig:sunset}}.
\end{equation}
Numerical results for the finite part are shown in
table~\ref{tab:sunset}
for two choices of parameters, comparing with values obtained through the dispersion
relation method of Ref.~\cite{disp}.
\begin{table}
\centering
\renewcommand{\arraystretch}{1.3}
\begin{tabular}{|l|l|l|}
\hline
$(p^2,m_1^2,m_2^2,m_3^2)$ & (1, 1, 4, 5)\hfill\anc & (16, 2, 2, 1)\hfill\anc \\
\hline
Result from MB repr.\ & 18.19558(2) & $17.5083(1)-0.1075(1)i$
\\[-.3ex] 
(with integration error) & & \\
\hline
Result from disp.\ rel.& 
 $18.19559$ &
 $17.50812\phantom{()}-0.10747i$\\
\hline
\end{tabular}
\mycaption{Numerical results for the finite part of the sunset diagram in
Fig.~\ref{fig:sunset}, obtained with the MB representation method described in
this report (second line) and the dispersion relation method of
Ref.~\cite{disp} (third line). Monte-Carlo integration errors for the last
quoted digit are given in
parentheses. The scale of dimensional regularization has been chosen $\mu=\,$1.
\label{tab:sunset}} 
\end{table}

\vspace{1ex}
The scalar five-propagator self-energy in Fig.~\ref{fig:master} can be cast into
a ten-dimensional MB representation, which can be reduced to integrals with
dimension up to six with
the help of the reduction formulas presented in the previous section.
Note that this integral is UV-finite, so that no terms from pole crossings
occur.
\begin{figure}[p]
\vspace{1ex}
\begin{center}
\rule{0mm}{0mm}%
\psline(-2,-1)(-1,-1)
\psline(1,-1)(2,-1)
\psline(0,0)(0,-2)
\pscircle(0,-1){1}
\psdot[dotscale=1.5](-1,-1)
\psdot[dotscale=1.5](1,-1)
\psdot[dotscale=1.5](0,0)
\psdot[dotscale=1.5](0,-2)
\rput[rt](-0.8,0){$m_1$}
\rput[lt](0.8,0){$m_4$}
\rput[rb](-0.8,-2){$m_2$}
\rput[lb](0.8,-2){$m_5$}
\rput[l](0.1,-1){$m_3$}
\rput[rb](-1.5,-1){$\displaystyle p\longrightarrow$}
\end{center}
\vspace{1.4cm}
\mycaption{Five-propagator two-loop diagram. As before, all propagators are
assumed to be scalars.
\label{fig:master}} 
\end{figure}

Owing to the relative large dimensionality of the integration, the numerical
integration is much slower than in the previous example. With about one hour
evaluation time on a single core of a Pentium IV computer with 2.4 GHz
a precision of 0.1\% to 1\% is obtained, depending on the values
of the masses and the momentum, see table~\ref{tab:master}.
They compare well with results based on the dispersion
relation method of Ref.~\cite{disp}.%
\begin{table}[p]
\centering
\renewcommand{\arraystretch}{1.3}
\begin{tabular}{|l|l|l|l|}
\hline
$(p^2,m_1^2,m_2^2,m_3^2,m_4^2,m_5^2)$ & ($-1$, 1, 1, 1, 1, 1)\hfill\anc 
 & (1, 3.7, 3.7, 1, 3.7, 3.7)\hfill\anc
 & (20, 1, 1, 1, 1, 1)\hfill\anc \\
\hline
Result from MB repr.\ & $-0.6809(4)$ & $-0.246(2)$ & $0.439(1)+0.259(7)i$
\\[-.3ex] 
(with integration error) & & & \\
\hline
Result from disp.\ rel.& 
 $-0.68088$ & $-0.25421$ & 
 $0.43952+0.25365i$\\
\hline
\end{tabular}
\mycaption{Numerical results for the self-energy diagram in
Fig.~\ref{fig:master}, obtained with the MB representation method described in
this report (second line) and the dispersion relation method of
Ref.~\cite{disp} (third line). Monte-Carlo integration errors for the last
quoted digit are given in
parentheses. The scale of dimensional regularization has been chosen $\mu=\,$1.
\label{tab:master}} 
\end{table}

\paragraph{Vertex topologies:} Figure~\ref{fig:vertex} shows a few examples for
two-loop vertex contributions with three or four different scales. The three
scalar diagrams lead to MB integrals with 9, 10, and 11 dimensions, respectively,
which can be reduced to five dimensions each using the reduction formulas in
eqs.~\eqref{mba}~ff. In table~\ref{tab:vertex}, numerical results are compared
with numbers obtained with the dispersion
relation method of Refs.~\cite{disp,ew2}.
\begin{figure}[p]
\vspace{1ex}
\begin{center}
\rule{0mm}{0mm}%
\psline[linestyle=dashed](-2,-1)(-1,-1)
\psline(-1,-1)(0,-0.5)(0,-1.5)(-1,-1)
\psline[linestyle=dashed](0,-0.5)(1,0)
\psline[linestyle=dashed](0,-1.5)(1,-2)
\psline(2,0)(1,0)(1,-2)(2,-2)
\psdot[dotscale=1](-1,-1)
\psdot[dotscale=1](0,-0.5)
\psdot[dotscale=1](0,-1.5)
\rput[bl](-2,-0.8){$Z$}
\rput[t](-0.5,-1.5){$b$}
\rput[b](-0.5,-0.5){$b$}
\rput[r](0,-1){$t\,$}
\rput[t](0.5,-0.4){$W$}
\rput[b](0.5,-1.5){$W$}
\rput[tr](2,-0.2){$b$}
\rput[br](2,-1.8){$b$}
\rput[l](1,-1){$\;t$}
\rput(0,-2.5){(a)}
\hspace{5.5cm}
\rule{0mm}{0mm}%
\psline[linestyle=dashed](-2,-1)(-1,-1)
\psline(-1,-1)(0,-0.5)(0,-1.5)(-1,-1)
\psline[linestyle=dashed](0,-0.5)(1,0)
\psline[linestyle=dashed](0,-1.5)(1,-2)
\psline(2,0)(1,0)(1,-2)(2,-2)
\psdot[dotscale=1](-1,-1)
\psdot[dotscale=1](0,-0.5)
\psdot[dotscale=1](0,-1.5)
\rput[bl](-2,-0.8){$Z$}
\rput[t](-0.5,-1.5){$t$}
\rput[b](-0.5,-0.5){$t$}
\rput[r](0,-1){$b\,$}
\rput[t](0.5,-0.4){$W$}
\rput[b](0.5,-1.5){$W$}
\rput[tr](2,-0.2){$b$}
\rput[br](2,-1.8){$b$}
\rput[l](1,-1){$\;t$}
\rput(0,-2.5){(b)}
\hspace{5.5cm}
\rule{0mm}{0mm}%
\psline[linestyle=dashed](-2,-1)(-1,-1)
\psline[linestyle=dashed](0,-0.5)(0,-1.5)
\psline[linestyle=dashed](1,0)(1,-2)
\psline(2,0)(1,0)(-1,-1)(1,-2)(2,-2)
\psdot[dotscale=1](-1,-1)
\psdot[dotscale=1](0,-0.5)
\psdot[dotscale=1](0,-1.5)
\rput[bl](-2,-0.8){$Z$}
\rput[t](-0.5,-1.5){$t$}
\rput[b](-0.5,-0.5){$t$}
\rput[l](0,-1){$\,H$}
\rput[tr](2,-0.2){$b$}
\rput[br](2,-1.8){$b$}
\rput[l](1,-1){$\;W$}
\rput(0,-2.5){(c)}
\end{center}
\vspace{2cm}
\mycaption{Planar two-loop vertex diagrams which appear as part of the two-loop
corrections to $Z\to b\bar{b}$. As before, all propagators are
assumed to be scalars, and external legs are constrained to be on-shell. 
\label{fig:vertex}} 
\end{figure}

\begin{table}[p]
\centering
\renewcommand{\arraystretch}{1.3}
\begin{tabular}{|l|l|l|l|}
\hline
Diagram & Fig.~\ref{fig:vertex}~(a) 
 & Fig.~\ref{fig:vertex}~(b)
 & Fig.~\ref{fig:vertex}~(c) \\
\hline
Result from MB repr.\ & $(2.24(2)+2.34(2)i)\times10^{-9}$ 
 & $1.04(1)\times10^{-9}$ & $3.78(3)\times10^{-10}$
\\[-.3ex] 
(with integration error) & & & \\
\hline
Result from disp.\ rel.& 
 $(2.241\phantom{()}+2.376i)\times10^{-9}$ & $1.036\times10^{-9}$ & 
 $3.805\times10^{-10}$\\
\hline
\end{tabular}
\mycaption{Numerical results for the vertex diagrams in
Fig.~\ref{fig:vertex}, obtained with the MB representation method described in
this report (second line) and the dispersion relation method of
Ref.~\cite{disp} (third line). Monte-Carlo integration errors for the last
quoted digit are given in
parentheses. The following values for the masses and regularization scale have
been used: $m_Z$=91, $m_W$=80, $m_t$=175, $m_H$=100, $\mu$=1.
\label{tab:vertex}} 
\end{table}

\paragraph{Box topologies:} This subsection will discuss
some two-loop box integrals
corresponding to typical topologies for next-to-next-to-leading order QCD
corrections to $pp \to t\bar{t}$, shown in Fig.~\ref{fig:tbox}. In the diagrams
massless (thin) lines correspond to light quark or gluon propagators,
while massive (thick) lines indicate top quark propagators.
\begin{figure}
\vspace{1ex}
\begin{center}
\rule{0mm}{0mm}%
\psline(-0.8,0)(0.8,0)
\psline(-0.8,-1.6)(0.8,-1.6)
\psline(0,0)(0,-1.6)
\psline(-1.7,0)(-0.8,0)(-0.8,-1.6)(-1.7,-1.6)
\psline[linewidth=1.6pt](1.7,0)(0.8,0)(0.8,-1.6)(1.7,-1.6)
\psdot[dotscale=1](-0.8,0)
\psdot[dotscale=1](0,0)
\psdot[dotscale=1](0.8,0)
\psdot[dotscale=1](-0.8,-1.6)
\psdot[dotscale=1](0,-1.6)
\psdot[dotscale=1](0.8,-1.6)
\rput(0,-2.1){(a)}
\hspace{4.2cm}
\rule{0mm}{0mm}%
\psline(-0.8,0)(0.8,0)
\psline(-0.8,-1.6)(0.8,-1.6)
\psline(-0.8,-0.8)(0.8,-0.8)
\psline(-1.7,0)(-0.8,0)(-0.8,-1.6)(-1.7,-1.6)
\psline[linewidth=1.6pt](1.7,0)(0.8,0)(0.8,-1.6)(1.7,-1.6)
\psdot[dotscale=1](-0.8,0)
\psdot[dotscale=1](-0.8,-0.8)
\psdot[dotscale=1](0.8,0)
\psdot[dotscale=1](-0.8,-1.6)
\psdot[dotscale=1](0.8,-0.8)
\psdot[dotscale=1](0.8,-1.6)
\rput(0,-2.1){(b)}
\hspace{4.2cm}
\rule{0mm}{0mm}%
\psline(0,0)(0,-1.6)
\psline(-1.7,0)(0.8,0)(0.8,-1.6)(-1.7,-1.6)
\psline[linewidth=1.6pt](1.7,0)(-0.8,0)(-0.8,-1.6)(1.7,-1.6)
\psdot[dotscale=1](-0.8,0)
\psdot[dotscale=1](0,0)
\psdot[dotscale=1](0.8,0)
\psdot[dotscale=1](-0.8,-1.6)
\psdot[dotscale=1](0,-1.6)
\psdot[dotscale=1](0.8,-1.6)
\rput(0,-2.1){(c)}
\hspace{4.2cm}
\rule{0mm}{0mm}%
\psline(-0.8,0)(-0.8,-1.6)
\psline(-1.7,0)(0.8,0)(0.8,-1.6)(-1.7,-1.6)
\psline[linewidth=1.6pt](1.7,0)(0,0)(0,-1.6)(1.7,-1.6)
\psdot[dotscale=1](-0.8,0)
\psdot[dotscale=1](0,0)
\psdot[dotscale=1](0.8,0)
\psdot[dotscale=1](-0.8,-1.6)
\psdot[dotscale=1](0,-1.6)
\psdot[dotscale=1](0.8,-1.6)
\rput(0,-2.1){(d)}
\end{center}
\vspace{1.6cm}
\mycaption{Some scalar master integrals relevant for two-loop QCD corrections
to $pp \to t\bar{t}$. Thick lines
indicate massive top-quark propagators, while thin lines are massless, and external
legs are constrained to be on-shell.
\label{fig:tbox}} 
\end{figure}

The master integrals corresponding to (a) and (b) in the figure lead to MB
integrals with dimension up to two and three, respectively, 
after expansion in $\varepsilon$. Therefore
they can be evaluated very efficiently with deterministic integration methods,
like Gaussian quadrature. These master integrals were computed previously in
Ref.~\cite{lhct2} with the method of differential equations. Comparing results
between the two methods\footnote{We are grateful to the authors of
Ref.~\cite{lhct2} for kindly providing numerical results of their calculation.},
agreement to more than nine significant digits has been obtained\footnote{Higher
precision could be achieved with more computing time.}. This comparison was
carried out both for an unphysical parameter point with Euclidean momentum
$s<0$, as well as for a physical parameter point with $s>(2m_t)^2$. For these
cases,  the evaluation of the MB integrals takes only several seconds to minutes
on a single-core Pentium IV computer.

To the best of our knowledge,
the two-loop integral corresponding to Fig.~\ref{fig:tbox}~(c) and (d) have not been
computed before. 
Fig.~\ref{fig:tbox}~(c) leads to a MB representation with a 13-dimensional integral,
which can be reduced to six dimensions with the help of the reduction formulas.
Numerical results for a few parameter points are shown in table~\ref{tab:tbox}.
Since the structure of the integral is rather complicated, the evaluation is
much slower than the previous examples. On a single-core Pentium IV computer
with 2.4 GHz, the numerical evaluation of one parameter point takes 4-5 hours to reach
the precision reported in the table.
\begin{table}
\centering
\renewcommand{\arraystretch}{1.4}
\begin{tabular}{|c|l|}
\hline
$(s,t,m_t^2)$ & Result from MB representation $\times 10^{15}$ (with integration error)  \\
\hline 
$(-100^2,\,-100^2,\,175^2)$
 & $-2.95397\varepsilon^{-2} + [50.15(1)+0.00(1)i]\varepsilon^{-1} -
 	[515.1(4)-0.0(4)i]$ \\
$\phantom{-}(400^2,\,-100^2,\,175^2)$
 & $[17.9514-13.6206i]\varepsilon^{-2} + [4720.1(5)-668.5(4)i]\varepsilon^{-1}$ 
 \\[-1ex] & 
$\qquad + [3651(13)-1481(13)i]$ \\
\hline
\end{tabular}
\mycaption{Numerical results for the box diagram in
Fig.~\ref{fig:tbox}~(c), obtained with the MB representation method.
Monte-Carlo integration errors are given in
parentheses. These numbers are based on $m_t=\,$175 for the top mass and
$\mu=\,$1 for the regularization scale.
\label{tab:tbox}} 
\end{table}

Fig.~\ref{fig:tbox}~(d) corresponds to a 9-dimensional MB integral, which
decomposes into integrals of dimension up to five after application of the
reduction formulas and expansion in $\varepsilon$. Table~\ref{tab:tbox2} shows
numerical results which have been obtained in about 10 hours evaluation time.
\begin{table}
\centering
\renewcommand{\arraystretch}{1.4}
\begin{tabular}{|c|l|}
\hline
$(s,t,m_t^2)$ & Result from MB representation $\times 10^{15}$ (with integration error)  \\
\hline 
$(-100^2,\,-100^2,\,175^2)$
 & $75.8804\varepsilon^{-3} 
 - [1141.55(5)+0.00(5)i]\varepsilon^{-2}$
\\[-1ex] & 
$\qquad + [13338.5(1.4)-0.5(1.4)i]\varepsilon^{-1} 
 -	[81900(530)+40(530)i]$ \\
$\phantom{-}(400^2,\,-100^2,\,175^2)$
 & $[4.17615+12.4156i]\varepsilon^{-3} 
 - [58.04(3)+272.30(3)i]\varepsilon^{-2}$
\\[-1ex] &
$\qquad + [252.3(9)+2932.3(9)i]\varepsilon^{-1} 
 +	[742(44)-20649(44)i]$ \\
\hline
\end{tabular}
\mycaption{Numerical results for the box diagram in
Fig.~\ref{fig:tbox}~(d), for the same conditions as in Tab.~\ref{tab:tbox}.
\label{tab:tbox2}} 
\end{table}


\section{Summary}
\label{sc:sum}

Mellin-Barnes representations are a powerful tool for the calculation of
multi-loop integrals. They have been shown to be very useful for systematically
extracting poles of the form $1/(4-D)$ in dimensional regularization. The
coefficients of these poles and the finite part can then be evaluated as
numerical integrals. However, for diagrams with massive propagators and physical
branch cuts, the integrands \emph{a priori} can be highly oscillatory, so that
standard numerical integration routines do not yield convergent results.

In this article, it is shown how the numerical convergence behavior can be
improved substantially by transforming the integration variables, deforming the
integration contours in the complex plane, and using analytical formulas to
reduce the dimensionality of the integration. The usefulness of those
modifications was demonstrated by calculating several examples of two-loop
self-energy, vertex and box topologies, some of which have been computed for the
first time.

Several previously known techniques permit much faster evaluation of relatively
simple multi-loop integrals with few independent masses and momenta than the
method presented here. However, Mellin-Barnes representations can be applied in
principle to any loop integral, irrespective of its complexity. With sufficient
computing time, arbitrarily precise results can be obtained.

In this work, tensor integrals with external uncontracted Lorentz indices were
not discussed separately. In fact, for most applications, it is convenient to
decompose the loop amplitudes into standard matrix elements or form factors, the
coefficients of which only contain scalar integrals. The integrands of these
scalar integrals can contain scalar products in the numerator involving on one
or more of the loop momenta. Such integrals can be evaluated straightforwardly
with Mellin-Barnes representations by expressing the scalar products as
propagators with negative powers.


\section*{Acknowledgements}

The authors would like to thank R.~Bonciani, A.~Ferroglia, and C.~Studerus for
useful discussions and providing numbers for comparison. This work has been
supported in part by the National Science Foundation under grant no.\
PHY-0854782.


\end{document}